\documentclass[10pt,A4paper,pra,twocolumn,amsmath]{revtex4}
\usepackage{graphicx}
\usepackage{bm}
\usepackage[section]{placeins}

\newcommand{\ket}[1]{|#1\rangle}             
\newcommand{\bra}[1]{\langle#1|}             
\newcommand{\braket}[2]{\langle#1|#2\rangle} 
\newcommand{\qo}[1]{``#1''}                  
\begin{document}
\title{Universal unitary gate for single-photon spinorbit four-dimensional states}
\author{Sergei Slussarenko$^{1}$, Ebrahim Karimi$^{1,2}$ , Bruno Piccirillo,$^{1,3}$ Lorenzo
Marrucci,$^{1,2}$ and Enrico Santamato$^{1,3}$}

\address{$^{1}$ Dipartimento di Scienze Fisiche, Universit\`{a} di
Napoli ``Federico II'', Compl.\ Univ.\ di Monte S. Angelo, 80126
Napoli, Italy\\
$^{2}$ CNR-INFM Coherentia, Compl.\ Univ.\ di Monte S. Angelo, 80126
Napoli, Italy\\
$^{3}$ Consorzio Nazionale Interuniversitario per le Scienze Fisiche
della Materia (CNISM), Napoli } \email{enrico.santamato@na.infn.it}
\begin{abstract}
The recently demonstrated possibility of entangling opposite values
of the orbital angular momentum (OAM) of a photon with its spin
enables the realization of nontrivial one-photon spinorbit
four-dimensional states for quantum information purposes. Hitherto,
however, an optical device able to perform arbitrary unitary
transformations on such spinorbit photon states has not been
proposed yet. In this work we show how to realize such a ``universal
unitary gate'' device, based only on existing optical technology,
and describe its operation. Besides the quantum information field,
the proposed device may find applications wherever an efficient and
convenient manipulation of the combined OAM and spin of light is
required.
\end{abstract}
\maketitle
\section{Introduction}\label{sec:intro}
Besides energy and linear momentum, photons may carry spin angular
momentum (SAM) and orbital angular momentum (OAM). SAM is associated
with the polarization and OAM with the transverse amplitude and
phase profile of the photon propagation
mode~\cite{frankearnold08,allen92}. The photon spin has been one of
the most exploited optical resources since the beginning of optics.
The photon OAM has received increasing attention in more recent
times as a resource in quantum and classical optics, since OAM
exists in an inherently multidimensional space. Information can thus
be encoded in higher dimensional
OAM-alphabets~\cite{molinaterriza02,molinaterriza07} to be used in
free-space communication systems~\cite{gibson04} or for increasing
the dimensionality of the working Hilbert space in quantum
communication systems~\cite{mair01}. In most applications
demonstrated thus far, however, only two opposite values of the OAM
were allowed, so as to encode one qubit into the OAM of a single
photon. A higher (four) dimensional (4D) space can then be still
realized by combining the OAM space with the usual spin, or
polarization, space, thus realizing what we call a \qo{spinorbit}
(SO) space.

In most multidimensional single photon optical schemes realized so
far, one qubit has been encoded in the photon spin
(\qo{polarization} qubit) and another qubit in the photon linear
momentum (\qo{direction} qubit) or even time-bin~\cite{marcikic02}. The two qubits can be also
entangled to each other, thus realizing full four-dimensional states (or ququarts) in
the 4D Hilbert space of the photon obtained by the tensor product of
the two qubit spaces. This approach requires however to split and
recombine the light beam in interferometric configurations which are
cumbersome and difficult to maintain aligned. It can be shown that
any unitary transformation in the polarization-direction 4D Hilbert
space can be realized by using a single interferometer and suitable
waveplates, thus creating an \emph{universal unitary gate} (UUG) in
that space~\cite{englert01}. Replacing the linear momentum with the
photon OAM, in quantum information encoding, has the advantage that
all processing can be made, in principle, along one beam, without
the need for interferometers.

Some manipulations of the photon OAM can be made by using Dove's
prisms~\cite{gonzales06}, transverse mode sorters~\cite{sasada03},
and cylindrical lens mode converters~\cite{beijersbergen93}, but no
UUG has been till now proposed for the photon OAM. However, even if
such OAM-UUG were available, its use together with a SAM-UUG would
still not reach the goal of a full manipulation of the ququart state
encoded in the photon, because these two UUGs would act on the spin
and OAM degrees of freedom separately and would not be able to
handle two-qubit entangling. Entangled pairs of qubits, i.e.,
unseparable ququarts, are central in most schemes that have been
proposed for quantum communications, quantum information processing
and secure key distribution.

In this work, we introduce a UUG acting in the 4D photon SO space
with OAM $m=\pm 2$, where the integer $m$ denotes the value of the
photon OAM along the optical beam axis, in units of $\hbar$. Hence,
we are restricting the infinite dimensional Hilbert space of the
photon OAM to the two-dimensional one $|m|=2$ (all our results are
easily generalized to an arbitrary $|m|$). Our SO-UUG is able to
handle arbitrary ququarts in this SO space, including those in which
the spin and orbital degrees of freedom of the photon are entangled.
Furthermore, our SO-UUG can be adjusted so as to obtain any unitary
transformation $U_{so}\in U(4)$ in the 4D photon SO-space, that will
turn any given input spinorbit state of the photon into any other
prescribed output spinorbit state. As a notable application, the
spectrum of any physical observable $\hat A$ in the SO-space can be
measured by means of a suitable SO-UUG adjusted so as to change the
eigenvectors of $\hat A$ into directly measurable basis states. An
important feature of the SO-UUG described here is that this device
can be inserted along the beam path without changing its direction
and without having recourse to interferometric schemes. Obviously,
our SO-UUG is much more complex than the SAM-UUG for photon spin
only, and it must have 16 adjustable parameters, i.e., as many as
the parameters of the $U(4)$ group. Nevertheless, the realization of
the SO-UUG is not beyond the possibilities of current technology.

This paper is organized as follows. In the next section we briefly
describe the UUG for photon spin only. In Sec.~\ref{sec:qplate}, we
describe the q-plate (QP), a recently invented device that may alter
the photon OAM and entangle it with its spin~\cite{marrucci06,
karimi09b,karimi09a,nagali09}. In Sec.~\ref{sec:qbox} we
introduce the so-called \qo{q-box} (QB), which is the main component
of our SO-UUG. The description of the whole SO-UUG is given in
Sec.~\ref{sec:SO-UUG}. In Sec.~\ref{sec:examples} some simple, yet
important, examples of ideal SO-UUGs are given. In
Sec.~\ref{sec:transverse} we discuss the problem of transverse-mode
cross-talk in a real SO-UUG, and, finally, in
Sec.~\ref{sec:conclusions} our conclusions are drawn.

\section{The spin unitary gate}\label{sec:SAM-UUG}
It is well known that the photon polarization may be easily
manipulated by inserting suitable birefringent plates along the
beam. A SAM-UUG for the photon spin is realized by an isotropic
retardation plate followed by a birefringent half-wave plate (HWP)
sandwiched between two birefringent quarter-wave plates (QWP) as shown in the inset of Fig.~\ref{fig:qbox}.
Changing the retardation $\delta$ of the isotropic plate and the
optical axis angles $\alpha,\beta,\gamma$ of the three birefringent
waveplates allows one to realize any unitary transformation
$\widehat{V}_s(\alpha,\beta,\gamma,\delta)\in U(2)$ in the photon
polarization space~\cite{swindell}. For example, Pauli's operator,
$\hat\sigma_x$, $\hat\sigma_y$, and $\hat\sigma_z$, in the circular
polarization basis of the spin space, may be realized by setting
$(\alpha,\beta,\gamma,\delta)=(0,\pi,\pi/2,\pi/2)\rightarrow\hat\sigma_x$,
$(\alpha,\beta,\gamma,\delta)=(0,\pi/4,0,\pi/2)\rightarrow\hat\sigma_y$,
$(\alpha,\beta,\gamma,\delta)=(0,-\pi/4,\pi/2,\pi/2)\rightarrow\hat\sigma_z$
(this choice is by no means unique). Denoting as
$\ket{1_{\hat{S}},m_{\hat{L}}}$ and $\ket{-1_{\hat{S}},m_{\hat{L}}}$
the photon states with left and right circular polarization and OAM
value $m$ respectively, the action of the  SAM-UUG is defined by its
action on the circular polarization basis states according to
\begin{equation}\label{eq:Vspin}
    (\ket{1_{\hat{S}},m_{\hat{L}}},\ket{-1_{\hat{S}},m_{\hat{L}}})\stackrel{\widehat{V}_s}
    {\rightarrow} (\ket{1_{\hat{S}},m_{\hat{L}}},\ket{-1_{\hat{S}},m_{\hat{L}}})\mathcal{V}_s
\end{equation}
where $\mathcal V_s$ is a $2\times2$ unitary matrix.

\section{The Q-plate}\label{sec:qplate}
Recently, a new device has been introduced, named q-plate (QP),
capable of producing entanglement between the spin and the OAM
degrees of freedom of a photon~\cite{marrucci06,karimi09b,karimi09a,nagali09}. The QP is
essentially a retardation wave-plate whose optical axis is aligned
non-homogeneously in the transverse plane in order to create a
topological charge $q$ in its orientation. An example of $q$-plate with topological charge $q=1$ is provided by the azimuthal alignement of the local optical axis, as shown in Fig.~\ref{fig:qbox}. QPs may be realized using
liquid crystals or other suitably patterned birefringent materials.
The optical behavior of the QP is characterized by its birefringent
retardation $\delta$ and its topological charge $q$. We say the QP
to be \qo{tuned} when its optical retardation is
$\pi$~\cite{karimi09b,karimi09a}. A tuned QP with charge $q=1$
performs a complete spin-to-OAM conversion of the photon state:
left- (right-) circularly polarized photons are converted into
right- (left-) handed photons and, simultaneously, the photon OAM
value $m$ is changed into $m+2$ ($m-2$), thus preserving the photon
total spinorbit angular momentum~\cite{marrucci06,karimi09b}. The
tuned QP acts as a polarization-controlled OAM mode shifter and can
induce entanglement between the spin and OAM degrees of
freedom~\footnote{Such entanglement cannot be achieved with
polarization-independent mode shifters such as fork holograms and
spatial light modulators, for example.}. The action of a tuned QP of
charge $q=1$ on the states $\ket{1_{\hat{S}},m_{\hat{L}}}$ and
$\ket{-1_{\hat{S}},m_{\hat{L}}}$ can be summarized as follows
\begin{equation}\label{eq:qplate}
    (\ket{1_{\hat{S}},m_{\hat{L}}},\ket{-1_{\hat{S}},m_{\hat{L}}})\stackrel{\widehat{QP}}
    {\rightarrow} (\ket{-1_{\hat{S}},(m+2)_{\hat{L}}},\ket{1_{\hat{S}},(m-2)_{\hat{L}}})
\end{equation}
In this way, a beam in an OAM state $m=\pm2$ is transformed into a
beam with $m=\pm4$ or $m=0$, depending on the polarization. The QP,
in fact, cannot be considered a gate in our SO-space with $m=\pm2$,
because it brings photons states out of this space. For this reason,
any gate acting in SO-space must contain a second cascaded QP which
restores the photon into the original OAM subspace.

An important issue that must be taken into account with all photon
OAM-shifters as QPs (as well as fork holograms) is the transverse
mode non-stationarity under free propagation, due to the beam radial
profile arising at the output of these devices. In fact, we should
have properly described our photon states by
$\ket{s_{\hat{S}},m_L}\ket{\varphi_m}$, where $\ket{\varphi_m}$ is
the radial state, whose evolution during propagation depends on the
OAM eigenvalue $m$. The transverse beam profile associated to the
radial state $\ket{\varphi_m}$ is given by the radial function
$\varphi_m(r)=\braket{r}{\varphi_m}$, where $\ket{r}$ denotes the
state of a photon localized at radial distance $r$ from the optical
axis. In general, the profile $\varphi_m(r)$ is a linear combination
of infinite Laguerre-Gauss (LG) radial profiles LG$_{pm}(r)$ with
given $m$ and different radial numbers $p$. During free propagation,
the Gouy phases of the different LG-mode components change relative
to each other, so that the total radial profile $\varphi_m(r)$
resulting from the superposition is not stationary. In most quantum
computation experiments, where OAM is not involved, the transverse
profiles are ignored, since they are the same for every state and
can be factorized out. We assumed such a factorization of radial
mode in writing Eq.~(\ref{eq:qplate}), which is justified when the
QP is optically thin and the radial mode is already factorized in
the input beam. But the right hand side of Eq.~\ref{eq:qplate} shows
that the beam emerging from the QP has two different $m$-components,
whose initially identical radial profile will change differently
during free propagation. As we shall see in a specific example, this
will give rise to some transverse \qo{cross-talk} between different
$m$-components of the beam reducing the fidelity of our SO-UUG.

\section{The Q-box}\label{sec:qbox}
\begin{figure}[t]
\includegraphics[scale=0.42]{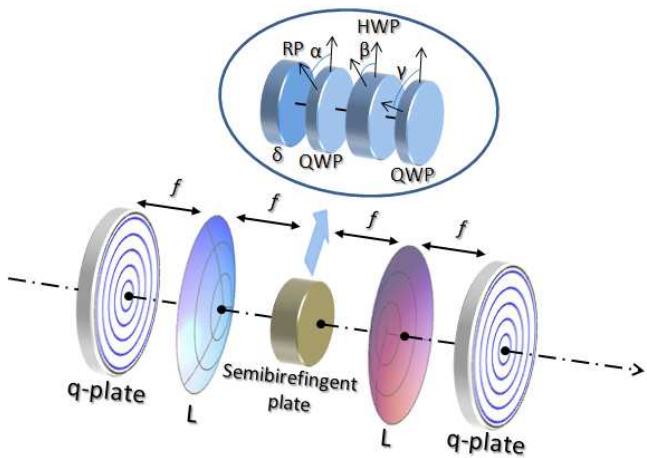}
\caption{\label{fig:qbox} (Color online) The scheme of the Q-box. The structure of
the birefringent plate is shown in the inset. QWP, HWP, RP are the
quarter-wave plate, half-wave plate and retardation plate
respectively. $\alpha$, $\beta$, $\gamma$ indicate rotation of the
QWP, HWP, QWP and $\delta$ is the retardation of the isotropic
plate. The birefringent plate realizes a SAM-UUG affecting the OAM $m=0$ part of the beam only and it is placed in the common focal plane of
the two lenses (L). The local optical axis of both q-plates is tangent to the
concentric circles so that $q=1$. The optical retardation both q-plate is $\lambda/2$.}
\end{figure}

The basic element of our SO-UUG is the device, here named q-box
(QB), that is shown in Fig.~\ref{fig:qbox}.
The QB is made by cascading two QPs, the first for spliting the OAM
$m=\pm 2$ components of the input beam into $m=0$ and $m=\pm4$
components, and the second to return them back into $m=\pm2$
components. Between the two QPs, a SAM-UUG is inserted which
performs an arbitrary unitary transformation
$\widehat{V}_s(\alpha,\beta,\gamma,\delta)$ as described in
Sec.~\ref{sec:SAM-UUG} on the photon spin only, and which acts
selectively on the OAM $m=0$ component of the beam, leaving the
$m=\pm 4$ components unaffected. This is made possible, because the
$m=0$ and $m=\pm 4$ components of the beam, initially superimposed
at the exit plane of the first QP, will spatially separate in their
radial pattern by free propagation. In fact, the $m=0$ part will
become concentrated in a spot at the beam center, and the $m=\pm4$
part will become distributed over a surrounding ring with zero
intensity at center (\qo{doughnut} beam shape). Having a good
separation between the $m=0$ and the $m=\pm4$ OAM components is
crucial for obtaining an efficient SO-UUG. This can be obtained in
several ways. In this work we consider a very simple sorting method,
based on a semibirefringent circular mask posed at the back focal
plane of the first lens shown in Fig.~\ref{fig:qbox}. The central
part of the mask is a birefringent disk of radius $R$, while the
surrounding corona is an isotropic medium. The central disk performs
the required unitary transformation $\widehat{V}_s$ on the photon
spin of the $m=0$ component. The radius $R$ of the active part of
the semibirefringent plate can be adjusted to mimimize the crosstalk
among the transverse radial modes. This minimization problem will be
discussed in Sec.\ref{sec:transverse} in a specific example. The
best spatial separation among radial modes occurs in the far field
or, equivalently, at the back focal plane of the first lens in
Fig.~\ref{fig:qbox}, where the Fourier transform of the field is
collected. The second lens in Fig.~\ref{fig:qbox} performs the
inverse Fourier transform thus restoring the input beam transverse
profile in the final output. To study the behavior of the QB in more
detail, it is convenient to use the following logical states in the
SO Hilbert space with OAM value $m=\pm 2$:
\begin{eqnarray}\label{eq:logical}
   \ket{00}&=&\ket{1_{\hat{S}},2_{\hat{L}}},\;\; \ket{01}=\ket{-1_{\hat{S}},-2_{\hat{L}}} \nonumber\\
   \ket{10}&=&\ket{-1_{\hat{S}},2_{\hat{L}}},\;\; \ket{11}=\ket{1_{\hat{S}},-2_{\hat{L}}}.
\end{eqnarray}
The action of any linear operator $\widehat{V}$ in the spinorbit
space is given by its action on the basis states, according to
\begin{equation}\label{eq:Vop}
    (\ket{00},\ket{01},\ket{10},\ket{11})\stackrel{\widehat{V}}{\rightarrow}
    (\ket{00},\ket{01},\ket{10},\ket{11})\mathcal{V}
\end{equation}
where $\mathcal V$ is a $4\times 4$ matrix. In particular, operators
as $\widehat{V}_s$ acting only on the photon spin are
mapped in spinorbit matrices having the general form
\begin{equation}\label{eq:Vmat}
   \mathcal V_s = \begin{pmatrix}
            v_{11} & 0 & v_{12} & 0 \\
            0 & v_{22} & 0 & v_{21} \\
            v_{21} & 0 & v_{22} & 0 \\
            0 & v_{12} & 0 & v_{11}
              \end{pmatrix}
\end{equation}
where $v_{ij}$ are the entries of the $2\times2$ matrix $\mathcal
V_s\in U(2)$ in Eq.~(\ref{eq:Vspin}). Similarly, operators
$\widehat{U}_o$ acting in the OAM subspace only are mapped in
spinorbit matrices having the general form
\begin{equation}\label{eq:Umat}
   \mathcal U_o = \begin{pmatrix}
         u_{11} & 0 & 0 & u_{12} \\
         0 & u_{22} & u_{21} & 0 \\
         0 & u_{12} & u_{11} & 0 \\
         u_{21} & 0 & 0 & u_{22}
                   \end{pmatrix}
\end{equation}
where $u_{ij}$ are the entries of the $2\times2$ matrix $\mathcal
U_o \in U(2)$. Assuming the first QP located in the front focal
plane of the first lens of the QB, in the back focal plane we have
the 2D-Fourier transform of the field exiting from the QP. The
2D-Fourier transform can be described by an operator
$\widehat{J}_{|m|}$ acting only on the radial part of the state:
viz.~ $\widehat{J}_{|m|}\ket{s_{\hat{S}},m_{\hat{L}},\varphi_m} =
\ket{s_{\hat{S}},m_{\hat{L}}}\widehat{J}_{|m|}\ket{\varphi_m}$. In
the basis of the localized states $\ket{r}$ in the Fourier plane
(lens back focal plane) the matrix elements of the operator
$\widehat{J}_{|m|}$ are given by
\begin{equation}\label{eq:J_lem}
   \bra{r'}\widehat{J}_{|m|}\ket{r}=\frac{1}{\lambda f}{J}_{|m|}\left(\frac{r' r}{\lambda f}\right)
\end{equation}
where ${J}_{|m|}(x)$ are the Bessel functions of integer order
$|m|$, $\lambda$ is the wavelength and $f$ is the lens focal length.
The operators $\widehat{J}_{|m|}$ are hermitian and unitary, so that
$\widehat{J}_{|m|}^2=1$. The action of a semi-birefringent plate (SBP)
with the birefringent disk of radius $R$ with corresponding unitary
spin operator $\widehat{V}_s$ is given by
\begin{eqnarray}\label{eq:semi}
&&\ket{s_{\hat{S}},m_{\hat{L}},\varphi_m}\stackrel{\widehat{SBP}}{\rightarrow}\nonumber\\
&&\widehat{V}_s\ket{s_{\hat{S}},m_{\hat{L}}}\widehat{U}(R)\ket{\varphi_m}+
   \ket{s_{\hat{S}},m_{\hat{L}}}(\widehat{I}-\widehat{U}(R))\ket{\varphi_m}
\end{eqnarray}
Where $\widehat{U}(R)$ is hermitian and idempotent operator that
acts only on the radial profile state, selecting the portion of the
beam at distance $r\leq R$ and $\widehat{I}$ is the identity operator.
In the basis of the localized states $\ket{r}$ the operator
$\widehat{U}(R)$ is represented by the Heaviside unit step function
$\Theta(R-r)$ with diagonal matrix elements
\begin{equation}\label{eq:step_matrix}
\bra{r'}\widehat{U}(R)\ket{r}=\Theta(R-r)\delta(r-r').
\end{equation}
The action of the QB on the logical basis states (\ref{eq:logical})
can now be  easily calculated
\begin{widetext}
\begin{eqnarray}\label{eq:qbox}
&&(\ket{00},\ket{01},\ket{10},\ket{11})\ket{\varphi_2}\stackrel{\widehat{Q}_b}{\rightarrow}(\ket{00},\ket{01},\ket{10},\ket{11}){\mathcal Q}_b\ket{\varphi_2}+\nonumber\\
&&((v_{11}-1)\ket{01}+(v_{22}-1)\ket{00}+v_{12}\ket{1_{\hat{S}},-6_{\hat{L}}}+v_{21}\ket{-1_{\hat{S}},6_{\hat{L}}})
\widehat{J}_4\widehat{U}(R)\widehat{J}_4\ket{\varphi_2}+\\
&&(\ket{10},\ket{11})({\mathcal I}-{\mathcal V_s})\widehat{J}_0
(\widehat{I}-\widehat{U}(R))\widehat{J}_0\ket{\varphi_m}\nonumber
\end{eqnarray}
\end{widetext}
where $v_{ij}$ are the entries of the $2\times2$ matrix $\mathcal
V_s$ of the birefringent disk, $\mathcal I$ is the 2D-identity
matrix and the ${\mathcal Q}_b$ is a $4\times4$ matrix given by
\begin{equation}\label{eq:qboxmat}
   \mathcal Q_b (\mathcal V_s) = \begin{pmatrix}
         1 & 0 & 0 &0 \\
         0 & 1 & 0 & 0 \\
         0 & 0 & v_{11} & v_{12} \\
         0 & 0 & v_{21} & v_{22}
                   \end{pmatrix}
\end{equation}
The last two terms in the left side of Eq.~(\ref{eq:qbox}) are the
crosstalk terms that appear due to the small overlap of $m=0$ and
$m=\pm4$ modes in the Fourier plane. The effects related to mode
cross-talk will be discussed below in a specific example. For the
moment,  we define the \textit{ideal} QB a QB where the transverse
mode cross-talk is negligible. The action of an ideal QB is then
given by
\begin{equation}\label{eq:qbox_short}
(\ket{00},\ket{01},\ket{10},\ket{11})\stackrel{\widehat{Q}_b}{\rightarrow}(\ket{00},\ket{01},\ket{10},\ket{11}){\mathcal Q_b}
\end{equation}
The matrix $\mathcal Q_b$ in Eq.~(\ref{eq:qboxmat}) has not the form
of the matrices in Eqs.~(\ref{eq:Vmat}) and (\ref{eq:Umat}), showing
that the QB acts on both the spin and OAM degree of freedom of the
photon simultaneously, producing entangling. Because the matrix
${\mathcal V}_s$ of the semi-birefringent plate is unitary, it is
evident from Eq.~(\ref{eq:qboxmat}) that also ${\mathcal Q}_b$ is
unitary and so is the action of the ideal QB. The main property of
the QB is that although the first QP takes photons out of our
initial SO logical space (\ref{eq:logical}), the operator of the
full ideal QB is well defined in this space, as shown by
Eq.~(\ref{eq:qbox_short}).

\section{The universal unitary gate}\label{sec:SO-UUG}
In this section, we show that the ideal QB defined by
Eq.~(\ref{eq:qbox_short}) can be used to make the required UUG in
our SO space. The QB itself does not provide the most general
unitary transformation of $U(4)$. We can see this just by
considering that ${\mathcal Q}_b(\mathcal V_s)$ (as ${\mathcal V}_s$
itself) depends on four real parameters only, while the most general
unitary operator $U_{so}\in U(4)$ has 16 free parameters. The UUG in
the spinorbit space is therefore a device more complex than the QB.
It is then remarkable that the sequence of
\begin{itemize}
\item   an ideal QB ${\mathcal Q}_b(\mathcal V_2)$
\item   a quarter-wave plate at angle 0
\item   an ideal QB ${\mathcal Q}_b(\mathcal V_R)$
\item   a half-wave plate at angle 0
\item   an ideal QB ${\mathcal Q}_b(\mathcal V_L)$
\item   a quarter-wave plate at angle 0
\item   an ideal QB ${\mathcal Q}_b(\mathcal V_1)$
\end{itemize}
provides the required UUG in the SO Hilbert space. Here $\mathcal
V_1,\mathcal V_R,\mathcal V_L,\mathcal V_2$ are the $2\times 2$
unitary matrices characterizing the semi-birefringent plates
inserted in each QB. We notice that all elements of the QB are
transparent so that many of them can be cascaded along the beam
direction maintaining optical losses at reasonable level. The SO-UUG
described above has the proper number of free parameters and it is
unitary. However, we should also demonstrate that it is universal,
i.e.\ that \textit{any} unitary matrix of $\mathcal U_{so}\in U(4)$
can be realized by the sequence above. The proof is based on the
results by Englert et al.~\cite{englert01}. In fact, a
straightforward calculation shows that the matrix $\mathcal U_{so}$
associated to our UUG has the block form
\begin{equation}\label{eq:S}
    \mathcal U_{so} =\begin{pmatrix}
                    \mathcal S_{LL} & \mathcal S_{LR} \\
                    \mathcal S_{RL} & \mathcal S_{RR}
                 \end{pmatrix}
\end{equation}
where the $2\times 2$ blocks are given by
\begin{eqnarray}\label{eq:Sentries}
    \mathcal S_{RR} &=& \frac{1}{2}\mathcal V_2(\mathcal V_R + \mathcal V_L)\mathcal V_1 \nonumber\\
    \mathcal S_{LL} &=& \frac{1}{2}(\mathcal V_R + \mathcal V_L) \nonumber \\
    \mathcal S_{RL} &=& -\frac{i}{2}\mathcal V_2(\mathcal V_R - \mathcal V_L) \nonumber \\
    \mathcal S_{LR} &=& \frac{i}{2}(\mathcal V_R - \mathcal V_L)\mathcal V_1.
\end{eqnarray}
Equations (\ref{eq:S}) and (\ref{eq:Sentries}) are identical to
Eqs.~(17) and (18) of Ref.~\cite{englert01}, where it is also shown
that for any given unitary matrix $\mathcal U_{so}\in U(4)$ one can
find four unitary matrices ${\mathcal V}_1,\mathcal V_R,\mathcal
V_L,\mathcal,\mathcal V_2 \in U(2)$ such that $\mathcal U_{so}$ has
the form (\ref{eq:S}) with (\ref{eq:Sentries}). It is worth noting
that all manipulations in our SO-UUG are made on the photon spin
degree of freedom, and that QPs are used to transfer the required
operations to the OAM degree of freedom, as in Ref.\
\cite{nagali09}.

One SO-UUG can perform any $U(4)$ transformation in the SO space
with $m=\pm2$. However, many nontrivial operations can be realized
with simpler devices. For example, a straightforward calculation
shows that one QB with a half-wave plate inside is enough to
implement the c-NOT operator on our logical state
$(\ket{00}\rightarrow\ket{00},\ket{01}\rightarrow\ket{01},\ket{10}\rightarrow\ket{11},\ket{11}\rightarrow\ket{10})$.
A more common set of logical states in the photon spinorbit space is
the set where the first bit corresponds to the photon spin and the
second bit to the photon OAM:
\begin{eqnarray}\label{eq:natural}
   \ket{0,0}&=&\ket{1_{\hat{S}},2_{\hat{L}}},\;\; \ket{0,1}=\ket{1_{\hat{S}},-2_{\hat{L}}} \nonumber\\
   \ket{1,0}&=&\ket{-1_{\hat{S}},2_{\hat{L}}},\;\; \ket{1,1}=\ket{-1_{\hat{S}},-2_{\hat{L}}}.
\end{eqnarray}
In this \qo{natural} set of basis states, the action of our SO-UUG
is different. For example, the matrix ${\mathcal Q}_n$ associated to
a single QB in the \qo{natural} states basis assumes the form
\begin{equation}\label{eq:qboxmatnat}
   \mathcal Q_n (\mathcal V_s) = \begin{pmatrix}
         1 & 0 & 0 &0 \\
         0 & v_{11} & v_{12} & 0 \\
         0 & v_{21} & v_{22} & 0 \\
         0 & 0 & 0 & 1
                   \end{pmatrix}
\end{equation}
in place of Eq.~(\ref{eq:qboxmat}). This way, the single QB which
realizes the c-NOT operation in the logical states basis
(\ref{eq:logical}), on the states (\ref{eq:natural}) performs the
swapping operation
$(\ket{00}\rightarrow\ket{00},\ket{01}\rightarrow\ket{10},\ket{10}\rightarrow\ket{01},\ket{11}\rightarrow\ket{11})$.
Since the choice (\ref{eq:natural}) of states is the most used in
the literature, we will use this \qo{natural} basis hereafter. Among
the gates that can be realized by the UUG, some are more important
than others. Here we list some of these useful gates indicating how
they can be implemented. It is worth noting, however, that the
implementation of a gate is not unique, and the gates presented here
can be realized, in some cases, in simpler ways.

\section{Examples}\label{sec:examples}
All examples shown in this section have been worked out in the
\qo{natural} basis of logical states (\ref{eq:natural}). We have already seen, that the
swapping gate is made of one QB with a HWP (i.e. ${\mathcal
V_s}=\sigma_x$) semibirefringent plate. The swapping gate is very useful,
because it allows one to transfer any unitary action made on the
spin qubit to the OAM qubit. For example, a gate mapping basis
states into equal-weight maximally entangled orthogonal
superpositions (called Hadamard gate), for the single-qubit in the
spin degree of freedom, is simply realized by a QWP oriented at
45\ensuremath{^\circ}. Insertion of the swapping gate after the QWP
yields a Hadamard gate acting onto the OAM qubit, leaving the spin
qubit unchanged for future manipulation. A Hadamard gate for general
ququarts can be realized with three QBs by setting ${\mathcal
V}_1=\sigma_y,{\mathcal V}_2=i\sigma_z,{\mathcal
V}_L=\sigma_x,{\mathcal V}_R=1$.

Another useful gate is the c-NOT gate. The c-NOT gate, realized by
UUG, can be either a spin-controlled or a OAM-controlled NOT gate.
The difference between them is whether the control bit is encoded in
the SAM qubit or in the OAM qubit, respectively. These two gates are
given by Eqs.~(\ref{eq:S}) and (\ref{eq:Sentries}) with ${\mathcal
V}_1=-i \sigma_x,{\mathcal V}_2=i\sigma_x,{\mathcal
V}_L=\sigma_z,{\mathcal V}_R=1$ for the spin c-NOT and ${\mathcal
V}_1=-i,{\mathcal V}_2=i,{\mathcal V}_L=\sigma_z,{\mathcal V}_R=1$
for the OAM c-NOT, respectively. These gates are very useful in most
quantum optics applications because of their universality.

A measurement of Bell states can be performed with a gate, which
transforms Bell state basis in the natural one, where each state can
be separated by common devices. The Bell's state measurement gate is
provided by ${\mathcal V}_1=1,{\mathcal V}_2=i\sigma_y,{\mathcal
V}_L={\mathcal V}_R=1/\sqrt{2}(1-i\sigma_y)$. This gate unties the
entanglement between the photon SAM and OAM allowing to measure the
two degrees of freedom separately.

\section{The radial modes crosstalk}\label{sec:transverse}
A final issue to be addressed to is the \qo{crosstalk} among radial
modes taking place in our device. This effect lowers, in general,
the overall quality of the QB (and hence of the SO-UUG) and must be
maintained at tolerable levels. As it can be seen from
Eq.~(\ref{eq:qbox}), when some overlap of $m=0$ and $m=\pm 4$ modes is
present, crosstalk terms appear in the QB.

\begin{figure}[htbp]
\includegraphics[width=6cm]{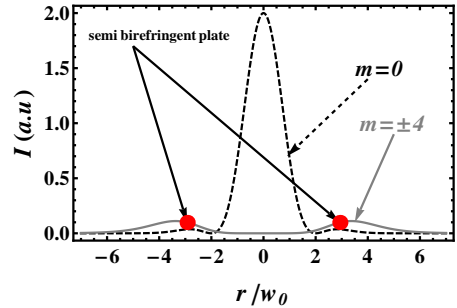}
\caption{\label{fig:qboxfield} (Color online) The transverse intensity profiles
produced by an incident LG$_{0,2}$ beam at the back focal plane of
the first lens in Fig.~\ref{fig:qbox}, where the semi-birefringent
plate is inserted. The two curves have been normalized to unit area,
for better showing the mode crosstalk.}
\end{figure}

We made an evaluation of the effects due to the radial cross-talk in
the case of the swapping gate (one QB with half-wave
semibirefringent plate), using the radial Laguerre-Gauss mode
LG$_{02}$ profile as input \cite{karimi07}. The radial field
profiles of the $m=0$ and $m=\pm4$ OAM components of the beam in the
back focal plane of the first lens are shown in
Fig.~\ref{fig:qboxfield}. We notice that, although the main part of
the power of the $m=0$ component is concentrated at the beam center,
where the $m=\pm4$ component vanishes, a small crosstalk is present
due to the Airy secondary maxima of the $m=0$ mode. As
indication of the quality of the QB as an element of the unitary
gate we used the \qo{fidelity} of the photon state
$\ket{\psi_{real}}$ produced by the swapping QB with respect to the
state $\ket{\psi_{ideal}}$ that an ideal QB would have produced. The
fidelity $F$ is here defined by
$F=|\braket{\psi_{real}}{\psi_{ideal}}|$. The input photon state and
the expected state from an ideal swapping gate are given by
\begin{equation}\label{eq:in_state}
\ket{\psi_{in}}=(a\ket{0,0}+b\ket{0,1}+c\ket{1,0}+d\ket{1,1})\ket{\mathrm LG_{02}}
\end{equation}
and
\begin{equation}\label{eq:out_state}
\ket{\psi_{ideal}}=(a\ket{0,0}+c\ket{0,1}+b\ket{1,0}+d\ket{1,1})\ket{\mathrm LG_{02}}
\end{equation}
respectively. The ideal QB given by Eq.~(\ref{eq:qboxmatnat}) with
$\mathcal V_s=\sigma_x$ has 100\% fidelity with respect to the
output state $\psi_{ideal}$. However, the action of a real swapping
gate on the input state (\ref{eq:in_state}) is described by
\begin{eqnarray}\label{eq:swap_real}
&&(a\ket{0,0}+b\ket{0,1}+c\ket{1,0}+d\ket{1,1})\ket{\mathrm LG_{02}}\stackrel{Swap}{\rightarrow}\nonumber\\
&&(a\ket{0,0}+c\ket{0,1}+b\ket{1,0}+d\ket{1,1})\ket{\mathrm LG_{02}}-\nonumber\\
&&[(c-b)(\ket{1,0}+\ket{0,1})\widehat{J}_0(\widehat{I}-\widehat{U}(R))\widehat{J}_0 - \\
&&(a(\ket{0,0}-\ket{-1_{\hat{S}},6_{\hat{L}}})+\nonumber\\
&&d(\ket{1,1}-\ket{1_{\hat{S}},-6_{\hat{L}}}))\widehat{J}_4 \widehat{U}(R) \widehat{J}_4] \ket{\mathrm LG_{02}}\nonumber
\end{eqnarray}
Assuming $b$ and $c$ to be real-valued, we can pose ($x^2=1-a^2-d^2$)
\begin{equation}\label{eq:subs}
b=x \cos\theta,~~c=x \sin\theta.
\end{equation}
The swapping gate fidelity function is then given by
\begin{eqnarray}\label{eq:fid}
F=|1-x^2(1-\sin2\theta)(1-\gamma_0(R))-(1-x^2)\gamma_4(R)|
\end{eqnarray}
where $\gamma_m(R)=\bra{\mathrm LG_{02}}\widehat{J}_m\widehat
U(R)\widehat{J}_m\ket{\mathrm LG_{02}}$ account for the radial mode
cross-talk. The parameters $\gamma_m(R)$ depend on the radius $R$ of
the active disk of the SBP inserted in the QB. In the case of the ideal QB, we have $\gamma_0(R)\rightarrow 1$, $\gamma_4(R)\rightarrow 0$, and $F\rightarrow 1$. We notice that the fidelity $F$
given by Eq.~(\ref{eq:fid}) depends on the input state of the
photon. The knowledge of the state to be swapped allows us to select
the radius $R$ of the birefringent disk so as to have maximum
fidelity (that can vary from 83\% to almost 100\%, in some cases). A
good compromise, however, can be obtained choosing $R=2.928$ (in
units of beam waist at the back focal plane), which gives us a flat
minimum fidelity of 83.7\% for \textit{any} input state. Among the
other ways to increase the fidelity of QB, one could, for example,
put a non-transparent belt around the birefringent disk in order to
absorb the photons in the mode-overlapping region. This method can
increase the fidelity of the QB up to 100\% at the price of
decreasing transmission efficiency of the device. Another method
could be to adjust the input beam profile or improving the design of the $q$-plate so as to minimize
crosstalk of $m=0$ and $m=\pm4$ modes. The use of higher-order OAM modes could also help. We finally note that the full spinorbit UUG has many reflecting surfaces (about 70) so that high quality multireflection coating of all surfaces is highly desirable\footnote{With 1\% reflection at each surface the total transmittance of the UUG is reduced to about 50\%. With 0.1\% reflection, the transmittance can be still larger than 90\%.}.

\section{Conclusions}\label{sec:conclusions}
In conclusion, we have introduced a new optical device, the QB, by
which universal unitary gates can be assembled to realize arbitrary
unitary transformations $U_{so}\in U(4)$ in the photon spinorbit
Hilbert space. All optical manipulations are made by SAM-UUG acting
on the photon spin alone, and their action is transferred to the OAM
degree of freedom by means of QPs. This introduces a practical
advantage, because the photon spin can be easily and rapidly
manipulated by using birefringent materials. Our SO-UUG is highly
transparent, which guarantees a high photon transmission efficiency,
and can be inserted along the beam path in as much the same way as
other optical components are inserted without changing the beam
direction. This allows us to perform nontrivial two-qubit handling
with optical assemblies having much less overall size and much more
stability against environmental noise than interferometers. We think
that QB and QB-based UUG could have an impact in all fields where
complete manipulation of the light OAM and polarization is needed
such as, for instance, in optical tweezers and traps, optical
communications, optical computing and fundamental quantum optics. 

\newpage
\end{document}